\documentclass[final]{ias2}
\usepackage{graphicx}
\usepackage{multirow}
\usepackage{array}

\usepackage{hyperref}
\begin{document}

\markboth{The origin of the solar magnetic cycle}{Arnab Rai Choudhuri}

\title{The origin of the solar magnetic cycle}

\author[ain]{Arnab Rai Choudhuri}
\email{arnab@physics.iisc.ernet.in}
\address[ain]{Department of Physics, Indian Institute of Science, Bangalore -- 560012, India}
\begin{abstract}
After summarizing the relevant observational data, we discuss how
a study of flux tube dynamics in the solar convection
zone helps us to understand the formation of
sunspots.  Then we introduce the flux transport dynamo model and assess
its success in modelling both the solar cycle and its departures from
strictly periodic behaviour.
\end{abstract}


\maketitle

\section{Introduction}

It is a great honour for me to give a plenary talk in the Chandra Centenary
Symposium.  As a graduate student of Gene Parker in the early 1980s, I had
the privilege of working for four years in an office about 4 or 5 doors down 
the corridor from Chandra's office.  Those of you who had visited University
of Chicago in those days may know that most of the  astronomy 
faculty and students
were in a building called Astronomy and Astrophysics Center.  The building
next to it---Laboratory for Astrophysics and Space Research---mainly housed
the large cosmic ray research group.  However, 
two of the most beautiful offices in that
building were given to two theoretical astrophysicists---Chandra 
and Parker.  Chandra stopped taking students
after a heart attack in the 1970s and there were no students working with him
when I was in Chicago. For a while I was the only theory student having a very
nice office in that building.  When I was attending the first AAS meeting of
my life, somebody asked me at the dinner 
table, ``How big is your theory group?'' I replied:
``It is a very small theory group with only three members.'' The next question was,
``Who are the members?''  I casually said: ``Oh, besides myself, the other two
members of our small theory group are Subrahmanyan Chandrasekhar and Eugene Parker.''

There was such an aura around Chandra that, like most other graduate students,
I was in an awe and always tried my best to keep away from him.  Since I was
the only other Indian in the building, Chandra seemed somewhat curious about
me and often asked Gene what I was doing.  Gene used to tell me that I should overcome
my fear of Chandra and should talk to him some time.  A few days after my first 
paper dealing with the solar dynamo problem appeared in ApJ [1], 
while walking along the corridor, I saw Chandra coming from the opposite direction.
Normally we would walk past each other as if we were strangers.  That day, to my
utter consternation, Chandra suddenly stopped when he came close to me and looked
straight into my eyes.  Then he said: ``I have seen your paper.  It is a nice
piece of work.'' Without giving me any time to recover from my dazed state or to
respond, Chandra immediately started walking away.  I should mention that I had
always been a great admirer of Chandra's style of writing.  Although that first
paper of mine presented a relatively unimportant calculation, it was deliberately
written in imitation of the Chandra style.  You may want to compare that paper
[1] with the famous S.\ Candlestickmaker paper! I have a hope that,
if Chandra were present here today, he would have taken some interest in the
subject of my presentation.

All of you know about the 11-year periodicity of the sunspot cycle.  There
also seems to be a 50-year periodicity in this field which you may not be
aware of! So let me begin by telling you about this 50-year periodicity.
It appears that major breakthroughs in this field take place approximately
at the intervals of 50 years. (1) A little more than 150 years ago, the
German amateur astronomer Schwabe [2] reported the first discovery of
the sunspot cycle.  (2) About 100 years ago, Hale [3] found the evidence
of Zeeman splitting in the spectra of sunspots, thereby concluding that
sunspots are regions of concentrated magnetic field.  It may be mentioned
that this was a momentous discovery in the history of physics because this
was the first time somebody found a conclusive evidence of large-scale
magnetic fields outside the Earth's environment.  Now we know that magnetic
fields are ubiquitous in the astronomical universe. With Hale's discovery, it
also became clear that the
sunspot cycle is essentially a magnetic cycle of the Sun.
(3) About 50 years ago, Parker [4] finally formulated the turbulent dynamo
theory, which still provides the starting point of our understanding how
magnetic fields arise in astronomical systems.

Even without a theoretical model of this 50-year periodicity, you should
be able to make a simple extrapolation and predict that another major 
breakthrough in this field should be taking place right now.  We are going
to argue that such a breakthrough is indeed happening at the present time.
Some of the earlier breakthroughs were achieved single-handedly by extraordinary
individuals like Hale and Parker.  Now we probably live in a less heroic
age.  The present breakthrough is a result of efforts due to many groups
around the world, in which our group in Bangalore also has made some 
contributions.  

\section{Some observational considerations}

\begin{figure}[t]
\centerline{\includegraphics[height=13cm,angle=+90]{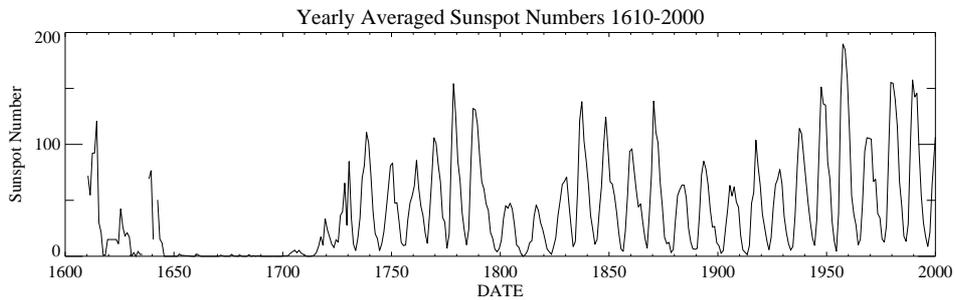}}
\caption{The yearly averaged number of sunspots plotted against time
for the period 1610--2000.}
\end{figure}

Let us begin by looking at Fig.~1, which plots the sunspot
number as a function of time from 1610. Galileo and some of
his contemporaries were the first scientists to study sunspots
systematically.  The initial entries in Fig.~1 are based on
their records.  Then, for nearly a century, sunspots were
rarely seen---a period known as the {\it Maunder minimum}.  Afterwards
the sunspot number has varied periodically with a rough period
of about 11 years, although we see a considerable amount of irregularity.
Some cycles are stronger than the average and some are weaker.
An intriguing question is whether we can predict the strength
of a cycle in advance.  Simple methods like expanding the last
few cycles in a Fourier series and continuing the series to predict
the next cycles have failed completely in the past. It is clearly
not a problem of merely extending a mathematical series and
we presumably need a proper understanding of what causes the
irregularities of the cycles if we hope to predict a future
cycle successfully.  When we discuss the causes of irregularities
in sunspot cycles in \S5, we shall address the question whether
our understanding of sunspot cycles 
at the present time is good enough to make
such predictions.

A few years after Schwabe's discovery of the sunspot cycle [2],
Carrington [5] noted that sunspots seemed to appear at lower
and lower latitudes with the progress of the solar cycle. 
It may be mentioned that 
individual sunspots live from a few days to a few
weeks. Most of the sunspots in the early phase of a solar
cycle are seen between $30^{\circ}$ and 
$40^{\circ}$.  As the cycle advances, new
sunspots are found at increasingly lower latitudes.  Then a 
fresh cycle begins with sunspots appearing again at high
latitudes.  Maunder [6] made the first graphical representation
of this.  In a time-latitude plot, the 
latitudes where sunspots were seen at
a particular time can be marked by black bars. Fig.~2 shows one
such plot.  The explanation of the grey-scale background will
be provided later. The sunspot
distribution in a time-latitude plot is often referred to as a
{\it butterfly diagram}, since the pattern (the regions marked in black
bars in Fig.~2) reminds one of butterflies. 

\begin{figure}[t]
\centering
\centerline{\includegraphics[height=6cm]{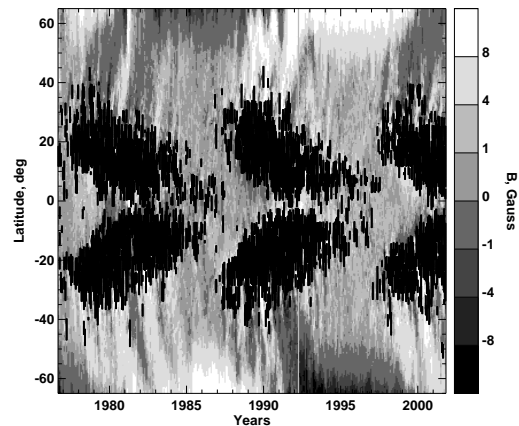}}
\caption{A `butterfly diagram' of sunspots,
with shades of grey showing the latitude-time distribution
of longitudinally averaged weak, diffuse magnetic field
($B$ is in Gauss).}
\end{figure}

We have mentioned Hale's discovery of magnetic fields in
sunspots [3]. A large sunspot has a typical magnetic
field of about 3000 G.
A few years later, Hale and his coworkers 
made another significant discovery [7].  
Often two large sunspots are seen side by side. Hale {\em et al}
[7] found that they invariably
have opposite polarities. Fig.~3 shows a magnetogram map
of the Sun in which white and black indicate respectively
regions of strong positive and negative polarities, grey
being put in regions where the magnetic field is below a
threshold.  A bipolar sunspot pair appears as a white patch and
a black patch side by side. You may note in Fig.~3 that the
right sunspots in the sunspot pairs in the northern hemisphere
are positive, whereas the right sunspots in the sunspot pairs 
in the southern hemisphere are negative.  This is the case for
a particular cycle.  In the next cycle, the polarity reverses.
The right sunspots in the northern hemisphere would become negative
in the next cycle and the right sunspots in the southern
hemisphere would become positive.  If we only look at the
sunspot number, we may think that the sunspot cycle has a
period of 11 years.  However, on taking account of the configuration
of the magnetic field, we realize that the Sun's magnetic cycle
has actually a period of 22 years. 

\begin{figure}[t]
\centerline{\includegraphics[height=8cm,width=8cm]{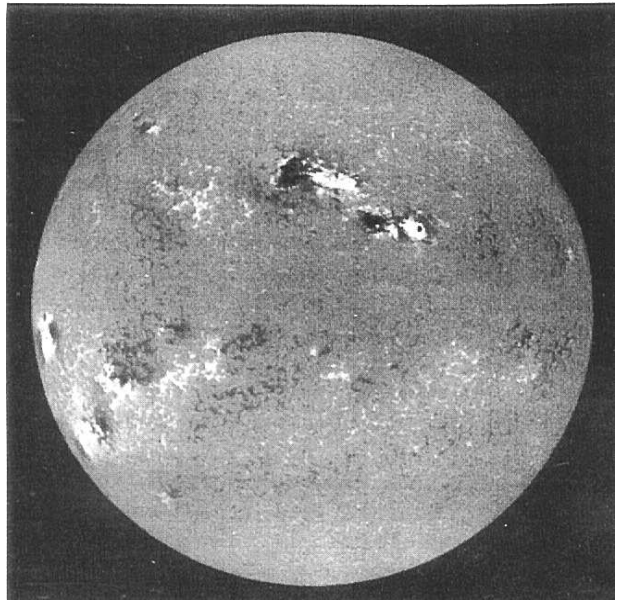}}
\caption{A magnetogram image of the full solar disk.  The 
regions with positive and negative magnetic polarities are
respectively shown in white and black, with grey indicating
regions where the magnetic field is weak.}
\end{figure}  

You may note another thing in Fig.~3. The line joining the centres of 
a bipolar sunspot pair is, on an average, nearly parallel to the solar
equator.  Hale's coworker Joy, however, noted that there is a
systematic tilt of this line with respect to the equator 
(the right sunspot in a pair appearing closer to the equator) and
that this tilt increases with latitude 
[7].  This result is
usually known as {\it Joy's law}.  The tilts, however, show
a considerable amount of scatter around the mean given by Joy's
law. As we shall see later, this law of tilts of sunspot
pairs plays a very important role in solar dynamo theory.

We shall present a detailed discussion in \S3 how the bipolar
sunspot pairs arise.  For the time being, let us just mention
that there has to be a strand of sub-surface magnetic field
which occasionally breaks out of the solar surface as shown
in Fig.~7b.  Then magnetic field lines would come out of one
sunspot (making its polarity positive) and would go down into
the other sunspot (making its polarity negative).  A look at
Fig.~3 suggests that there must be a sub-surface magnetic field
with field lines going from the right to the left in the northern
hemisphere and there must be an oppositely directed magnetic field
in the southern hemisphere.  Such a magnetic field in the azimuthal
direction is called a toroidal field.  This seems to be the
dominant component of the magnetic field in the Sun. In contrast,
the magnetic field of the Earth seems to be of poloidal nature.

In his seminal paper on the turbulent dynamo,  Parker [4]
proposed that the sunspot cycle is produced by an oscillation between toroidal and
poloidal components of the Sun's magnetic
field, just as we see an oscillation between kinetic and potential energies
in a simple harmonic oscillator. This was a truly extraordinary suggestion because almost
nothing was known about the Sun's poloidal field at that time. Babcock 
and Babcock
[8] were the first to detect the weak poloidal field having a strength of about 10 G
near the Sun's poles. Over the last few years, there is increasing evidence that the
field outside sunspots is actually not weak and diffuse, 
but concentrated in intermittent flux
concentrations [9].  Only in low-resolution magnetograms in which these flux
concentrations are not resolved, the field appears weak and diffuse. This seems
to be the case even in the polar regions [10].  However, we
shall not get into a more detailed discussion of this point here. 
Only from mid-1970s, we have systematic data of the Sun's polar
fields. Fig.~4 shows the polar fields of the Sun plotted as a function of time
(N and S indicating north and south poles), with the
sunspot number plotted below. It is clear that the sunspot number, which is a proxy
of the toroidal field, is maximum at a time when the polar field is nearly zero. On the
other hand, the polar field is strongest when the sunspot number is nearly zero. This
clearly shows an oscillation between the toroidal and poloidal components, as envisaged
by Parker [4]. The theoretical reason behind this oscillation
will be discussed in \S4.

\begin{figure}
\center
\includegraphics[width=8cm]{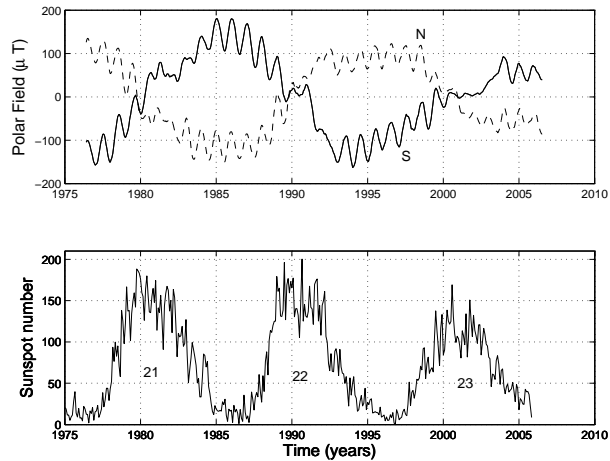}
\caption{The polar field of the Sun as a function of time
(on the basis of the Wilcox Solar Observatory data)
with the sunspot number shown below. The top panel, with
N and S representing north and south poles, is adapted
from [51]. The yearly modulations in the measurements
of the polar fields is due to the fact that the Sun's polar
axis is slightly inclined to the orbital plane of the
Earth's revolution around the Sun.}
\end{figure}

Let us make some more remarks on the poloidal field of the
Sun. It was found that there were large unipolar patches of 
diffuse magnetic field on
the solar surface which migrated poleward 
[11].  Even when
averaged over longitude, one finds predominantly one polarity
in a belt of latitude which drifts poleward 
[12]. The reversal
of polar field, which occurs at the time of the sunspot
maximum [13], presumably takes place when sufficient field of
opposite polarity has been brought to the poles.  
Along with the butterfly diagram of sunspots,
Fig.~2 also shows the distribution
of the longitude-averaged poloidal field in a time-latitude plot.  
The various shades of grey 
indicate values of the longitude-averaged
poloidal field. While the sunspots appear at lower and lower
latitudes with the progress of the solar cycle, the poloidal
field migrates poleward.  The reason behind the poleward 
migration of the poloidal field is a meridional circulation
in the Sun which involves a flow of gas at the surface from
the equatorial region to the polar region, having an amplitude
of about 20 m s$^{-1}$ [14].  
The poloidal field is carried poleward by this meridional circulation. 

It may be noted that all the observations discussed above pertain
to magnetic fields at the Sun's surface.  
We have no direct information about 
magnetic fields underneath the Sun's surface.  In dynamo theory,
we need to study the interactions between the magnetic fields
and velocity fields.  So let us now look at the nature of
the velocity fields of the Sun.  

Stellar structure models
suggest that the energy produced by nuclear reactions at the
centre of the Sun is transported outward by radiative transfer
to a radius of about $0.7 R_{\odot}$ (where $R_{\odot}$
is the solar radius). However, 
from about $0.7 R_{\odot}$ to $R_{\odot}$, energy is transported
by convection.  This region is called the {\em convection zone},
within which the plasma is in a turbulent state with hot
gas going up and cold gas coming down. The turbulent diffusivity
of the convection zone is the main source of diffusion in the
dynamo problem.  We have already pointed out that there is a
meridional circulation which is poleward near the solar surface
at the top of the convection zone.  This meridional circulation
is supposed to be driven by the turbulent stresses in the convection
zone, though our theoretical understanding of this subject is
rather limited at the present time. It is generally believed
that the meridional circulation at the bottom of the convection
zone has to go from the polar region to the equatorial region
in order to conserve mass, although we do not have a direct
evidence for it yet.  This meridional circulation plays a tremendously
important role in current dynamo models, as we shall see later.

\begin{figure}
\centering
\includegraphics[width=7cm]{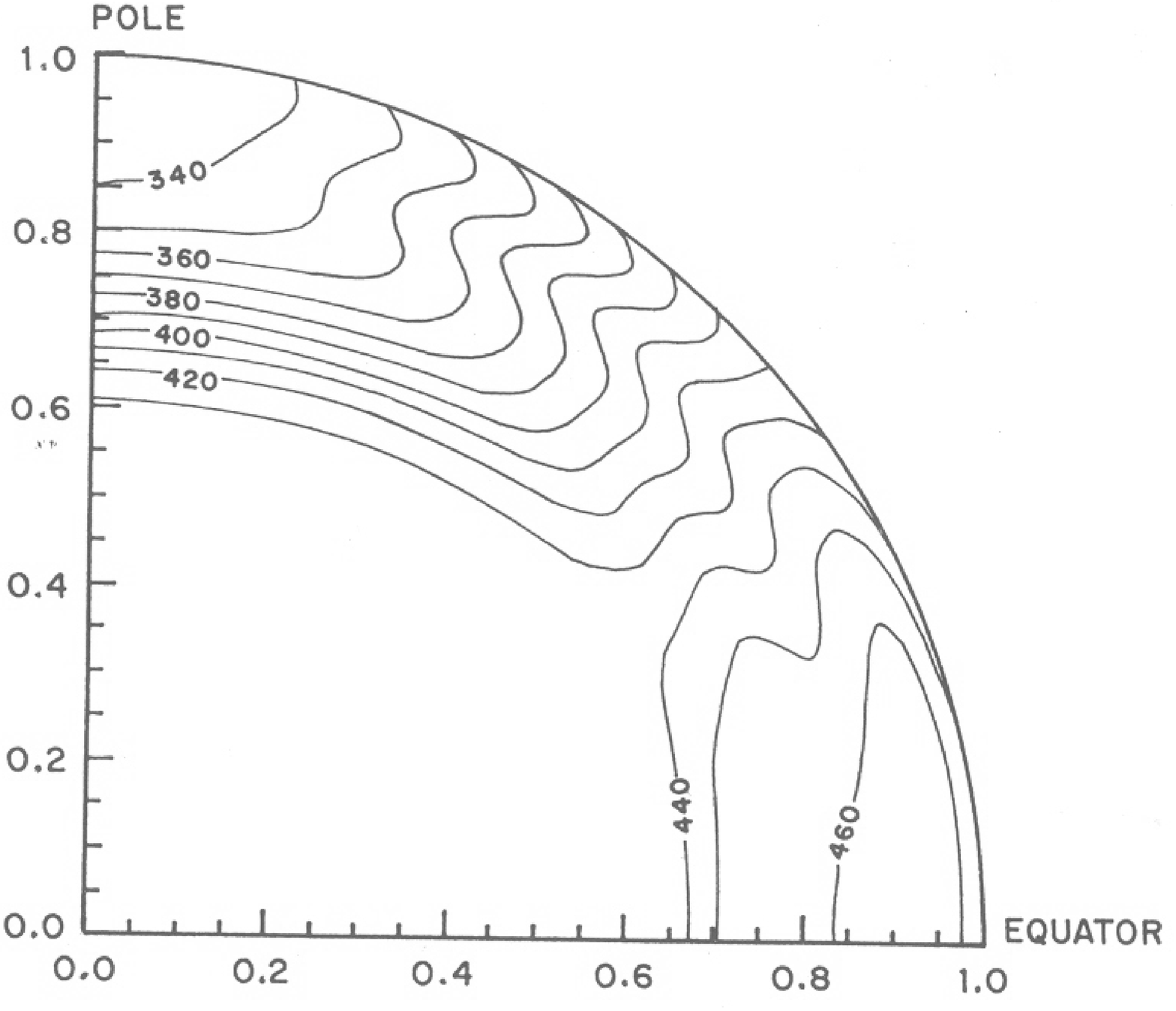}
\caption{The contours of constant angular velocity inside
the Sun, as obtained by helioseismology. The contours are
marked with rotation frequency in nHz.  It may be noted that
frequencies of 340 nHz and 450 nHz correspond respectively
to rotation periods of 34.0 days and 25.7 days.  Courtesy:
J. Christensen-Dalsgaard and M. J. Thomson.} 
\end{figure}

We finally come to what is probably the most important part of
the Sun's velocity field for us---the differential rotation.  Unlike
the Earth which rotates like a solid body, the Sun has the angular
velocity varying over it.  It has been known for a long time that
the angular velocity near the Sun's equator is faster than that
at the Sun's polar regions.  In the early years of dynamo research,
theorists used to make various assumptions about the distribution
of angular velocity in the Sun's interior.  An amazing development
of the last few decades has been helioseismology---the study of
the oscillations of the Sun.  These oscillations have allowed us
to probe various properties of the solar interior.  One of the
most extraordinary outcomes of helioseismology is that solar physicists
have been able to construct a map of angular velocity distribution
in the interior of the Sun (see, for example,
[15]).  A version of this map is shown in
Fig.~5. It is seen that there is strong differential rotation
(i.e.\ a strong gradient of angular velocity) at the bottom of
the solar convection zone.  This relatively thin layer of concentrated
differential rotation is called the {\em tachocline}.     

We have now come to an end of our discussion of what we know
about the magnetic and the velocity fields of the Sun.  The aim of
solar dynamo theory is the following.  Given our knowledge of
the velocity fields of the Sun, we need to study the interactions
between the velocity and magnetic fields in the Sun's interior
such that all the surface
observations of magnetic fields are properly explained---a fairly
daunting problem, of which the full solution is still a distant
dream.

\def\vb{{\bf v}}
\def\Bb{{\bf B}}
\def\ol{\overline}
\def\ec{\cal{E}}
\def\pa{\partial}
\def\vf{{\bf v}}
\def\Bf{{\bf B}}

\section{Formation of sunspots}

All our theoretical considerations are based on magnetohydrodynamics
or MHD.  An introduction to its basic concepts can be found in  
Chs.~14--16 of [16] or Ch.~8 of [17]. Let
us begin by mentioning some concepts of MHD which we shall be
using repeatedly.  We know that a magnetic field has a pressure
$B^2 /2 \mu$ associated with it, along with a tension along the
field lines. The other result which is going to be of central
importance to us is the theorem due to Alfv\'en [18] that, when
the magnetic Reynolds number is sufficiently high, magnetic fields
are frozen in the plasma and get carried by the velocity fields
of the plasma.  Because of the high magnetic Reynolds number in
the Sun, we expect this theorem to hold---at least approximately. 

Since energy is transported by convection in the layers below
the Sun's surface, sunspots are basically regions of concentrated
magnetic field sitting in a convecting fluid. To understand why
the magnetic field remains concentrated in structures like
sunspots instead of spreading out more evenly, we need to study
the interaction of the magnetic field with the convection in the
plasma.  This subject is known as {\it magnetoconvection}. 
The linear theory of convection in the presence of a
vertical magnetic field was studied by Chandrasekhar [19]. 
The nonlinear evolution of the system, however, can only be
found from numerical simulations pioneered by Weiss [20].  
Since the tension of magnetic field lines opposes convection, it
was found that space gets separated into two kinds of regions.
In certain regions, magnetic field is excluded and vigorous
convection takes place.  In other regions, magnetic field gets
concentrated, and the tension of magnetic field lines suppresses
convection in those regions.  Sunspots are presumably such regions
where magnetic field is piled up by surrounding convection.  Since
heat transport is inhibited there due to the suppression of convection,
sunspots look darker than the surrounding regions. 
Although we have no direct information about the state of the
magnetic field under the Sun's surface, it is expected that the
interactions with convection would keep the magnetic field
concentrated in bundles of field lines throughout the solar
convection zone.  Such a concentrated bundle of magnetic field
lines is called a {\it flux tube}. 

\begin{figure}[t]
\centerline{\includegraphics[height=5cm,width=10cm]{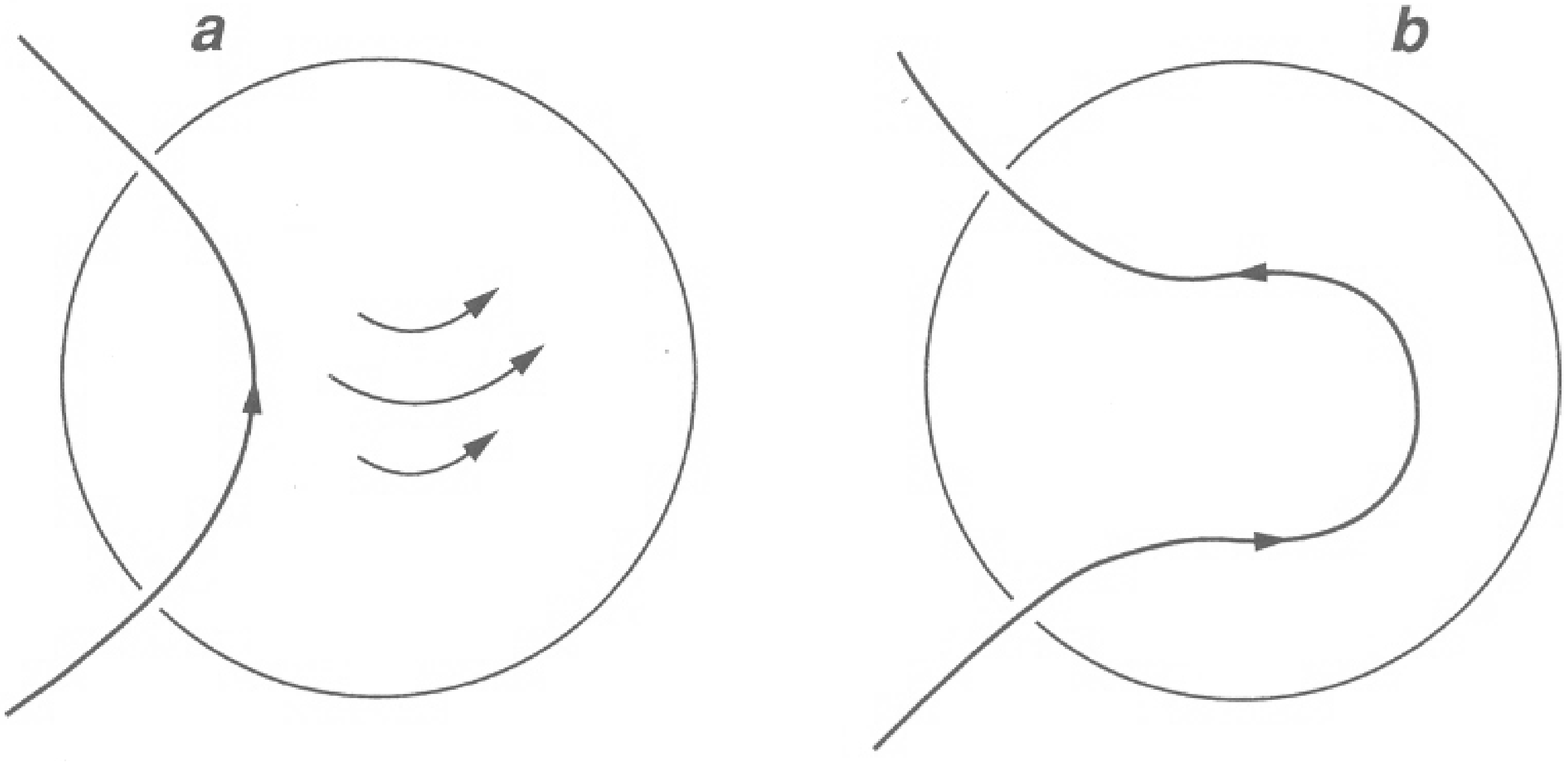}}
\caption{The production of a strong toroidal magnetic field
underneath the Sun's surface. {\bf a.} An initial poloidal field
line. {\bf b.} A sketch of the field line after it has been stretched
by the faster rotation near the equatorial region.}
\end{figure} 

One important consequence of Alfv\'en's theorem of flux freezing
for the Sun is the following.  If there is any poloidal field line
going through the Sun, differential rotation will drag it out
to produce a toroidal field, as shown in Fig.~6. The production
of the toroidal field is expected to be strongest in the tachocline
at the bottom of the convection zone where the gradient of angular
velocity is concentrated,
as seen in Fig.~5. Due to interactions with convection
there, the toroidal field should exist in the form of horizontal
flux tubes.  If a part of such a flux tube rises up and
pierces the solar surface as shown in Fig.~7b, then we expect
to have two sunspots with opposite polarities at the same
latitude.  But how can a configuration like Fig.~7b arise?
The answer to this question was provided by Parker [21] through
his idea of magnetic buoyancy.  We need to have a pressure
balance across the surface of a flux tube.  Since the magnetic
field inside the flux tube has a pressure 
$B^2/2 \mu$, the interior pressure is a sum of this pressure
and the gas pressure $p_{\rm in}$. On the other hand, the
only pressure outside is the gas pressure $p_{\rm out}$.  Hence
we must have
$$p_{\rm out} = p_{\rm in} + \frac{B^2}{2 \mu} \eqno(1)$$
to maintain pressure balance across the surface of a flux tube.  It
follows that
$$p_{\rm in} \leq p_{\rm out}, \eqno(2) $$
which often, though not always, implies that the density inside
the flux tube is less than the surrounding density.  If this happens
in a part of the flux tube, then that part becomes buoyant and
rises against the gravitational field to produce the configuration
of Fig.~7b starting from Fig.~7a. It is seen in Fig.~6b
that the toroidal fields in the two hemispheres are in the opposite
directions.  If parts of these toroidal fields rise in the two
hemispheres to produce the bipolar sunspot pairs, we have a natural
explanation why the sunspot pairs should have the opposite polarity
in the two hemispheres as seen in the magnetogram map of
Fig.~3.

\begin{figure}[t]
\centerline{\includegraphics[height=5cm,width=10cm]{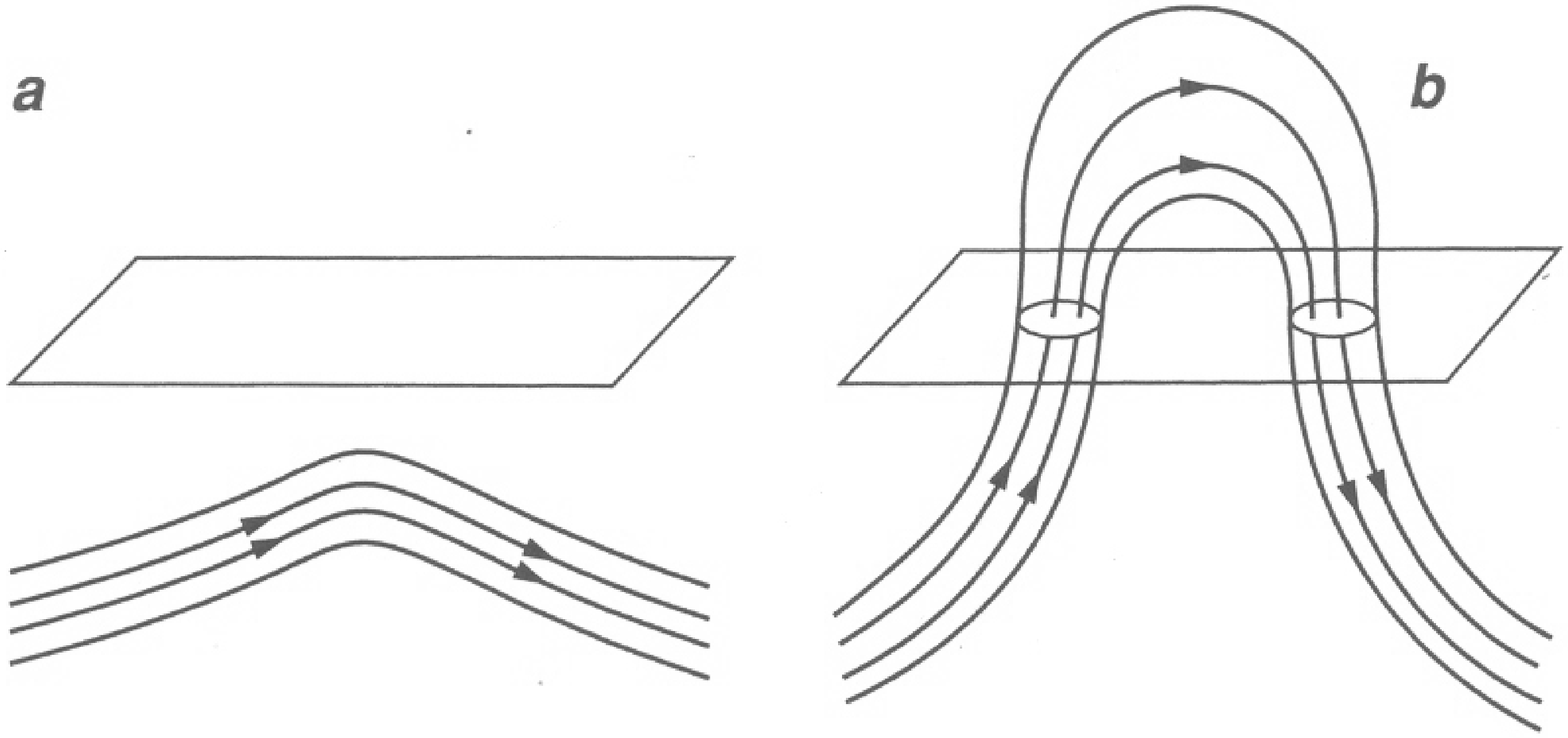}}
\caption{Magnetic buoyancy of a flux tube. {\bf a.} A nearly
horizontal flux tube under the solar surface. {\bf b.} The flux
tube after its upper part has risen through the solar surface.}
\end{figure}

It can be shown that
magnetic buoyancy is particularly destabilizing in the interior of
the convection zone, where convective instability and magnetic
buoyancy reinforce each other.  On the other hand, if a region is stable
against convection, then magnetic buoyancy can be partially
suppressed there (see, for example, \S8.8 in [22]). Since the
toroidal flux tube is produced at the bottom of the convection zone,
we may expect some parts of it to come into the convection zone and become
buoyant, whereas other parts may remain underneath the bottom of the
convection zone and stay anchored there due to the suppression of
magnetic buoyancy. A part of the flux tube coming within the convection
zone is expected to rise and eventually reach the solar surface to 
form sunspots, as sketched in Fig.~7. In order to model the formation
of bipolar sunspots, we have to study the dynamics of flux tubes
rising through the convection zone due to magnetic buoyancy.

The best way to study this problem is to treat it as an initial-value problem.
First, an initial configuration with a magnetic flux ring at the 
bottom of the convection zone, having a part coming inside
the convection zone, is specified, and then its subsequent
evolution is studied numerically. The evolution depends on the
strength of magnetic buoyancy, which is in turn determined by the
value of the magnetic field.  We shall give arguments in \S4
why most of the dynamo theorists till the early 1990s believed
that the magnetic energy density should be in equipartition with the
kinetic energy density of convection, i.e.
$$\frac{B^2}{2 \mu} \approx \frac{1}{2} \rho v^2. \eqno(3)$$
This suggests $B \approx 10^4$ G on the basis of standard models
of the convection zone.  If we use full MHD equations to study the
evolution of the flux tube, then the calculations become extremely
complicated.  However, if the radius of cross-section of the flux tube is
smaller than the various scale heights, then it is possible to derive
an equation for flux tube dynamics from the MHD equations [23, 
24]. Even this flux tube equation is a sufficiently
complicated nonlinear equation and has to be solved numerically.  
The evolution of such magnetic flux tubes due to magnetic buoyancy 
(starting from the bottom of the convection zone) was studied by Choudhuri 
and Gilman [25] and Choudhuri [26].  It was found that the Coriolis
force due to the Sun's rotation plays a much more important role
in this problem than what anybody suspected before.  If the initial
magnetic field is taken to have a strength of around $10^4$ G as
suggested by (3), then
the flux tubes move parallel to the rotation axis and emerge at
very high latitudes rather than at latitudes where sunspots are
seen.  Only if the initial magnetic field is taken as strong as
$10^5$ G, then magnetic buoyancy is strong enough to overpower
the Coriolis force and the magnetic flux tubes can rise radially
to emerge at low latitudes. 

\begin{figure}[t]
\centerline{\includegraphics[height=7cm,width=9cm]{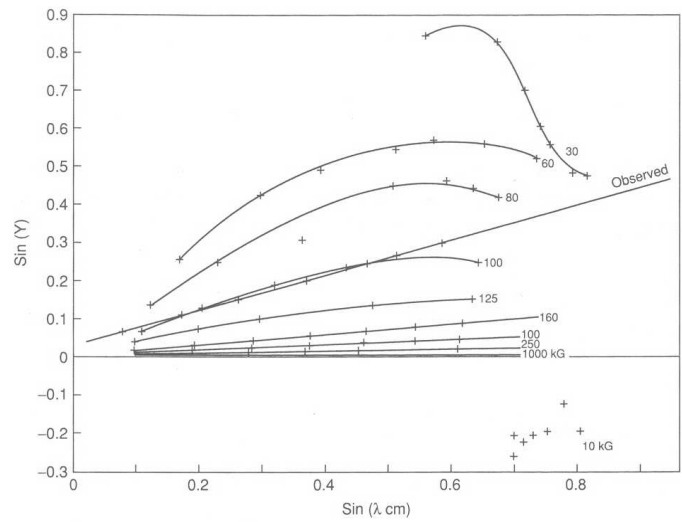}}
\caption{Plots of sin(tilt) against sin(latitude) theoretically
obtained for different initial values of magnetic field indicated
in kG. The observational data indicated by the straight line
fits the theoretical curve for initial magnetic field 100 kG
(i.e. $10^5$ G).  Reproduced from D'Silva and Choudhuri [27].}
\end{figure}

D'Silva and Choudhuri [27] extended these calculations to look
at the tilts of emerging bipolar regions at the surface. These tilts
are also produced by the action of the Coriolis force on the
rising flux tube.  Fig.~8
taken from [27] shows the 
observational tilt vs.\ latitude
plot of bipolar sunspots (i.e.\ Joy's law) along with the theoretical
plots obtained by assuming different values of the initial magnetic
field.  It is clearly seen that theory fits observations only if
the initial magnetic field is about $10^5$ G.  If the magnetic
field is much stronger, then the Coriolis force is unable to
produce much tilt.  On the other hand, flux tubes with weaker
magnetic fields are diverted to high latitudes. Apart from providing
the first quantitative explanation of Joy's law nearly three-quarters
of a century after its discovery, D'Silva and Choudhuri [27] put the first
stringent limit on the value of the toroidal magnetic field at the
bottom of the convection zone. Several other authors [28, 29] 
soon performed similar calculations and confirmed
the result.  Initially some efforts were made to explore whether
flux tubes with magnetic field given by (3) could satisfy various
observational constraints by invoking extra effects [30, 31].
However, the evidence kept mounting that the magnetic field
at the bottom of the convection zone is indeed much stronger than
the equipartition value given by (3). 

We already mentioned that the tilts of
active regions have a large amount of scatter around the mean given
by Joy's law.   In fact, it is found that active regions
often emerge with initial tilts inconsistent with Joy's law and then
the tilts change in the next few days to come closer
to values given by Joy's law [32].  Longcope and Choudhuri [33] 
have argued that the vigorous convective turbulence in the upper layers
of the convection zone exerts a random force on the tops of the rising
flux loops, causing the scatter around the Joy's law, and then the tilt
of the flux tube relaxes to the appropriate value after the emergence 
of the top of the tube through the solar surface when 
the top is no longer kicked by convective turbulence.

\section{Modelling the cycles from flux transport dynamo}

If we begin by assuming the Sun to have a poloidal field as shown
in Fig.~6a, we saw that various properties of sunspot pairs
can be explained.  The differential rotation would stretch this poloidal field
to produce the toroidal field, the interaction with convection
would lead to toroidal flux tubes and then magnetic buoyancy
would make these flux tubes rise to produce the bipolar sunspots.
However, if there is no mechanism to replenish the poloidal
field, then it would decay away and ultimately the whole
process outlined here would stop.  We now turn to the question
how the poloidal field is produced.  We invoke a mechanism
first proposed by Babcock [34] and Leighton [35].
The name of Leighton should be known to most physicists
as the second author of the celebrated {\em Feynman Lectures}
[36].

Let us now explain what this Babcock--Leighton mechanism is.
We pointed out in \S2 that bipolar sunspots have tilts increasing
with latitude, in accordance with Joy's law. Then we discussed in \S3
how this law was explained by D'Silva and Choudhuri [27] by
considering the action of the Coriolis force on rising flux tubes.
Now, a typical sunspot lives for a few days and the magnetic field
of the sunspot diffuses in the surrounding region by turbulent
diffusion after its decay.  When a tilted bipolar sunspot pair
with the right spot nearer the equator and the left spot
at a higher latitude decays, the polarity of the right sunspot
gets more diffused in the lower latitudes and the polarity of the
left sunspot gets more diffused in the higher latitudes. 
Take a look at Fig.~3 to visualize this process. This process
essentially gives rise to a poloidal field at the solar surface.
Since sunspots form from the toroidal field due to magnetic buoyancy,
a tilted bipolar sunspot pair can be viewed as a conduit through
which a part of the toroidal field ultimately gets transformed
into the poloidal field.  The tilted sunspot pair forms from
the toroidal and we get the poloidal field after its decay.
This is the basic idea of poloidal
field generation proposed by Babcock [34] and Leighton [35].

It may be noted that this Babcock--Leighton mechanism is somewhat
different from the original proposal of Parker [4], which
was elaborated further by Steenbeck, Krause \& R\"adler [37].
According to this original proposal, the turbulence in the
convection zone would involve helical motions due to the Coriolis
force and the toroidal field would be
twisted by this helical turbulence to produce the poloidal
field. However, this process,
often known as the $\alpha$-effect, can occur only if the maximum
value of the toroidal magnetic field is such that the magnetic
energy density does not exceed the kinetic energy of turbulence,
as indicated by (3).  As we already pointed out, flux tube
simulations for modelling sunspot formation suggest that the
toroidal field is about one order of magnitude stronger (about
$10^5$ G) compared to what we get from (3) (about $10^4$ G).
If the toroidal field is so strong, then the $\alpha$-effect
as originally envisaged by Parker [4] cannot work and the
Babcock--Leighton mechanism seems to be the likely mechanism
by which the poloidal field is produced.

\begin{figure}
\center
\includegraphics[width=6cm]{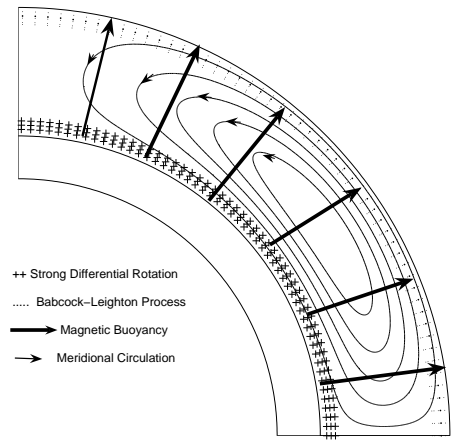}
  \caption{A cartoon explaining how the solar dynamo works within
the convection zone.}
\end{figure}

Fig.~9 is a cartoon encapsulating how the solar dynamo operates.
If you understand this cartoon, then you would have got the
central point of this presentation!
The toroidal field is produced in the tachocline by the differential
rotation stretching out the poloidal field. Then this toroidal
field rises due to magnetic buoyancy to produce bipolar sunspots
at the solar surface, where the poloidal field is generated by
the Babcock--Leighton mechanism from these bipolar sunspots.
The poloidal field so generated is carried by the meridional
circulation first to the polar region and then underneath the
surface to the tachocline to be stretched by the differential
rotation---thus completing the cycle. The likely streamlines
of meridional circulation are indicated in Fig.~9. This type 
of dynamo model in which the meridional circulation plays
a crucial role is
called a {\it flux transport dynamo}.  

Most of the dynamo theorists
at the present time believe that the solar dynamo operates in this
way.  Wang, Sheeley and Nash [38] proposed the idea of the flux
transport dynamo.  Choudhuri, Sch\"ussler and Dikpati [39] and
Durney [40] were the first to construct two-dimensional models
of the flux transport dynamo to demonstrate that such a dynamo
really does work. Initially it was thought that this type of
dynamo model would not work due to a technical reason.  There is
a rule, known as the {\it Parker-Yoshimura sign rule} 
[4, 41], which suggests that the type of dynamo outlined
in Fig.~9 would produce a poleward dynamo wave.  In other words,
it was feared that such a theoretical model would suggest that sunspots should
appear at higher and higher latitudes with the progress of the
sunspot cycle rather than at lower and lower latitudes.  Choudhuri,
Sch\"ussler and Dikpati [39] solved this puzzle by demonstrating
that a sufficiently strong meridional circulation can override
the Parker-Yoshimura sign rule and make the dynamo wave propagate
equatorward.  This paved the way for the subsequent growth of
the flux transport dynamo model.

So far in this presentation I have avoided getting into equations.
For those who wish to see the equations, I now show the central
equations of the flux transport dynamo theory.
In spherical coordinates, we write the
magnetic field as
$$
{\bf B} = B (r, \theta) {\bf e}_{\phi} + \nabla \times [ A
(r, \theta) {\bf e}_{\phi}],
\eqno(4)$$
where $B (r, \theta)$ is the toroidal component and $A (r, \theta)$
gives the poloidal component. We can write the velocity field
as $\vb + r \sin \theta \, \Omega (r, \theta) {\bf e}_{
\phi}$, where 
$\Omega (r, \theta)$ is the angular velocity in the interior of the
Sun and $\vb$ is the velocity of meridional circulation having components
in $r$ and $\theta$ directions.  Then the main equations telling us
how the poloidal and the toroidal fields evolve with time are
$$
\frac{\pa A}{\pa t} + \frac{1}{s}(\vf.\nabla)(s A)
= \lambda_T \left( \nabla^2 - \frac{1}{s^2} \right) A + \alpha B,
\eqno(5)$$
$$ \frac{\pa B}{\pa t} 
+ \frac{1}{r} \left[ \frac{\pa}{\pa r}
(r v_r B) + \frac{\pa}{\pa \theta}(v_{\theta} B) \right]
= \lambda_T \left( \nabla^2 - \frac{1}{s^2} \right) B 
+ s(\Bf_p.\nabla)\Omega + \frac{1}{r}\frac{d\lambda_T}{dr}
\frac{\partial}{\partial{r}}(r B), \eqno(6)$$
where $s = r \sin \theta$ and $\lambda_T$ is the turbulent
diffusivity inside the convection zone.   We should point out that
(5) and (6) are
mean field equations obtained by averaging over the turbulence
in the convection zone and describe the mean behaviour of
the average magnetic field. Since (5) and (6)
are coupled partial differential equations,
nothing much can be done
analytically. Our research group in IISc 
Bangalore has developed a numerical
code {\it Surya} for studying the flux transport dynamo problem
by solving (5) and (6).  I can
send the code {\it Surya} and a detailed guide for using it 
to anybody who sends a request to my e-mail address arnab@physics.iisc.ernet.in.

\begin{figure}
\center
\includegraphics[width=7cm]{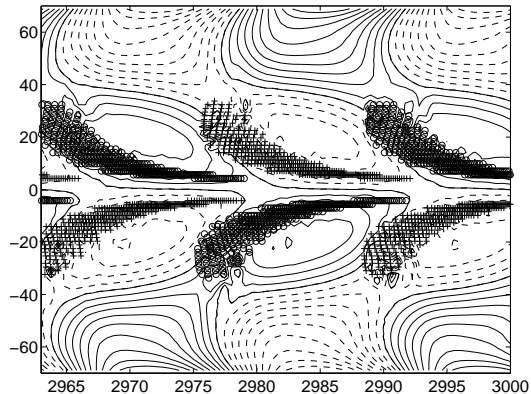}
  \caption{A theoretical butterfly diagram
of sunspots superposed on contours of constant $B_r$ at the solar surface
in a time-latitude plot. This figure is taken from Chatterjee, Nandy and
Choudhuri [43].}
\end{figure}

Some of the first results obtained with {\it Surya} were presented by Nandy and
Choudhuri [42] and Chatterjee, Nandy and Choudhuri [43]. I may mention
that a modified version of {\it Surya} has even been used to study the
evolution of magnetic fields in neutron stars [44, 45].  
Fig.~10 shows a theoretical butterfly diagram
of sunspots, superposed on contours in the time-latitude plot of the 
poloidal field on the solar surface.  This theoretical figure obtained
by the code {\it Surya} has to be
compared with the corresponding observational figure given in Fig.~2.
Given the fact that this was one of the first efforts of reproducing this
observational figure from a theoretical model, hopefully most readers will
agree that the match between theory and observations is not too bad.

The original flux transport dynamo model of Choudhuri, Sch\"ussler
and Dikpati [39] was developed at a time when Mausumi Dikpati was
my PhD student.  Afterwards she went to work in HAO Boulder and
the parent model led to two offsprings: a high diffusivity model
and a low diffusivity model.  The diffusion times across the
convection zone in these two models
are of the order of 5 years and 200 years respectively.  The high
diffusivity model has been developed in IISc Bangalore by me
and my successive PhD students
(Choudhuri, Nandy, Chatterjee, Jiang, Karak), whereas the low diffusivity
model has been developed by Dikpati and her co-workers in HAO (Dikpati,
Charbonneau, Gilman, de Toma).  The differences between these models
have been systematically studied by Jiang, Chatterjee and Choudhuri
[46] and Yeates, Nandy and Mckay [47].  Both these models are
capable of producing oscillatory solutions resembling solar
cycles.  However, when we try to study the variabilities of the
cycles, the two models give completely different results.  We need
to introduce fluctuations to cause variabilities in the cycles.
In the high diffusivity model, fluctuations spread all over the
convection zone in about 5 years.  On the other hand, in the low
diffusivity model, fluctuations essentially remain frozen during
the cycle period.  Thus the behaviours of the two models are totally
different on introducing fluctuations. It may be mentioned that simple
mixing length arguments suggest a reasonably high turbulent diffusivity
(see p.\ 629 of [22]) consistent with what is used in the high
diffusivity model of the IISc Bangalore group. 

\section{Irregularities of solar cycles and prospects for predicting
future cycles}

Before coming to the question of what causes the irregularities
of solar cycles, we take another look at the plot of polar
fields in Fig.~4.  The polar field at
the end of cycle~22 was weaker than the polar field in the previous
sunspot minimum.  We see that this weaker polar field was followed by
the cycle~23 which was weaker than the previous cycle.  Does this
mean that there is a correlation between the polar field during
a sunspot minimum and the next sunspot cycle?  In the left
panel of Fig.~11, we plot
the polar field in the sunspot minimum along the horizontal axis and
the strength of the next cycle along the vertical axis.  Although
there are only 3 data points so far, they lie so close to a
straight line that one is tempted to conclude that there is
a real correlation. There is a joke that astrophysicists
often do statistics with one data point, whereas here
we have three! On the other hand, the right panel
of Fig.~11, which has
the cycle strength along the horizontal axis and the polar field
at the end of that cycle along the vertical axis, has points
which are scattered around.  Choudhuri, Chatterjee and Jiang
[48] proposed the following to explain these observations.
While an oscillation between toroidal and poloidal components
takes place, the system gets random kicks at the epochs indicated
in Fig.~12.  Then the poloidal field and the next toroidal field
should be correlated, as suggested by the left panel of
Fig.~11.  On the other
hand, the random kick ensures that the toroidal field is not
strongly correlated with the poloidal field coming after it, as seen
in the right panel of Fig.~11.  

\begin{figure}
\begin{minipage}[c]{0.55\textwidth}
\vspace{0.1cm}
\includegraphics[height=5cm,width=5cm]{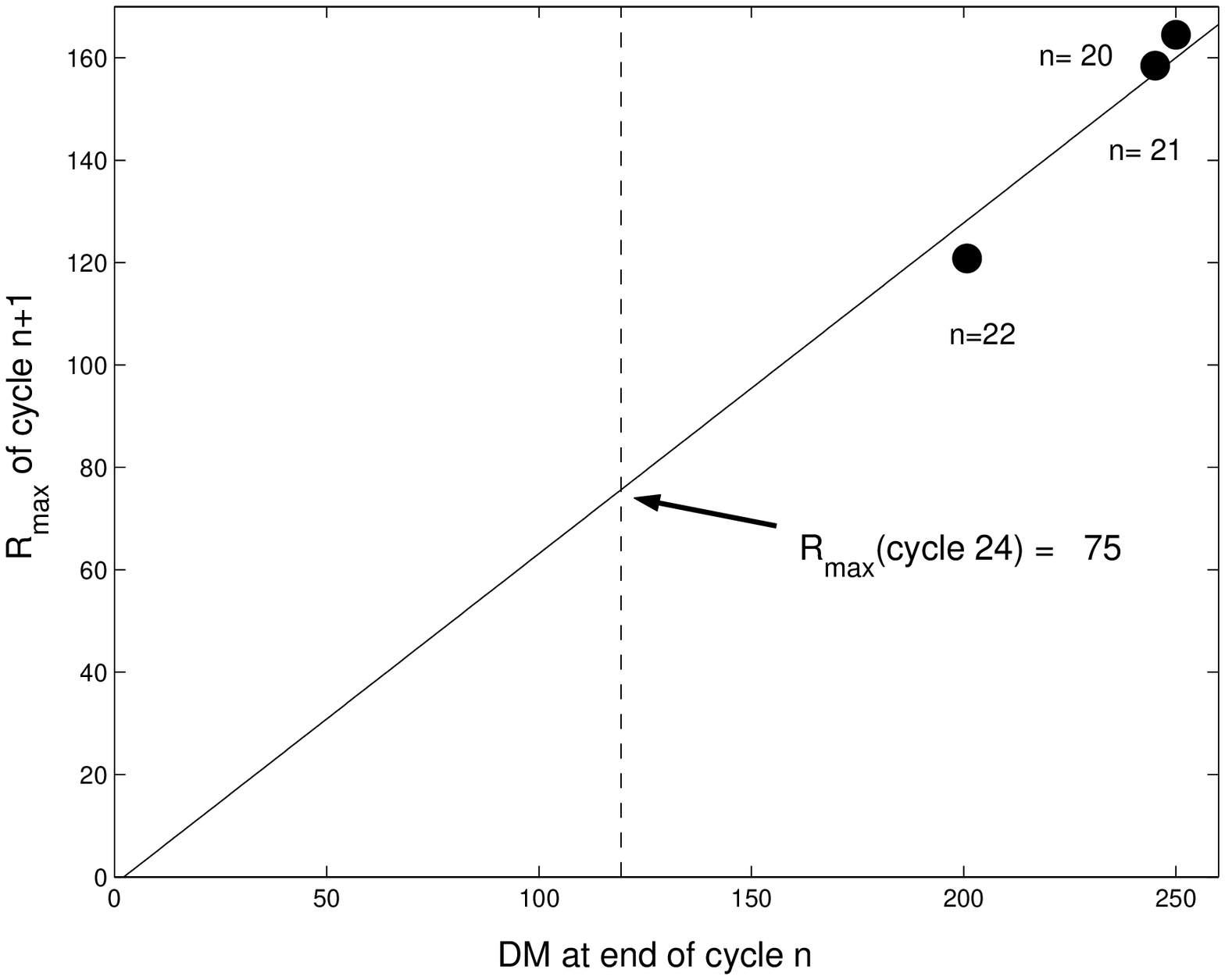}
\end{minipage}%
\hspace{0.3cm}
\begin{minipage}[c]{0.6\textwidth}
\includegraphics[height=4.5cm,width=5cm]{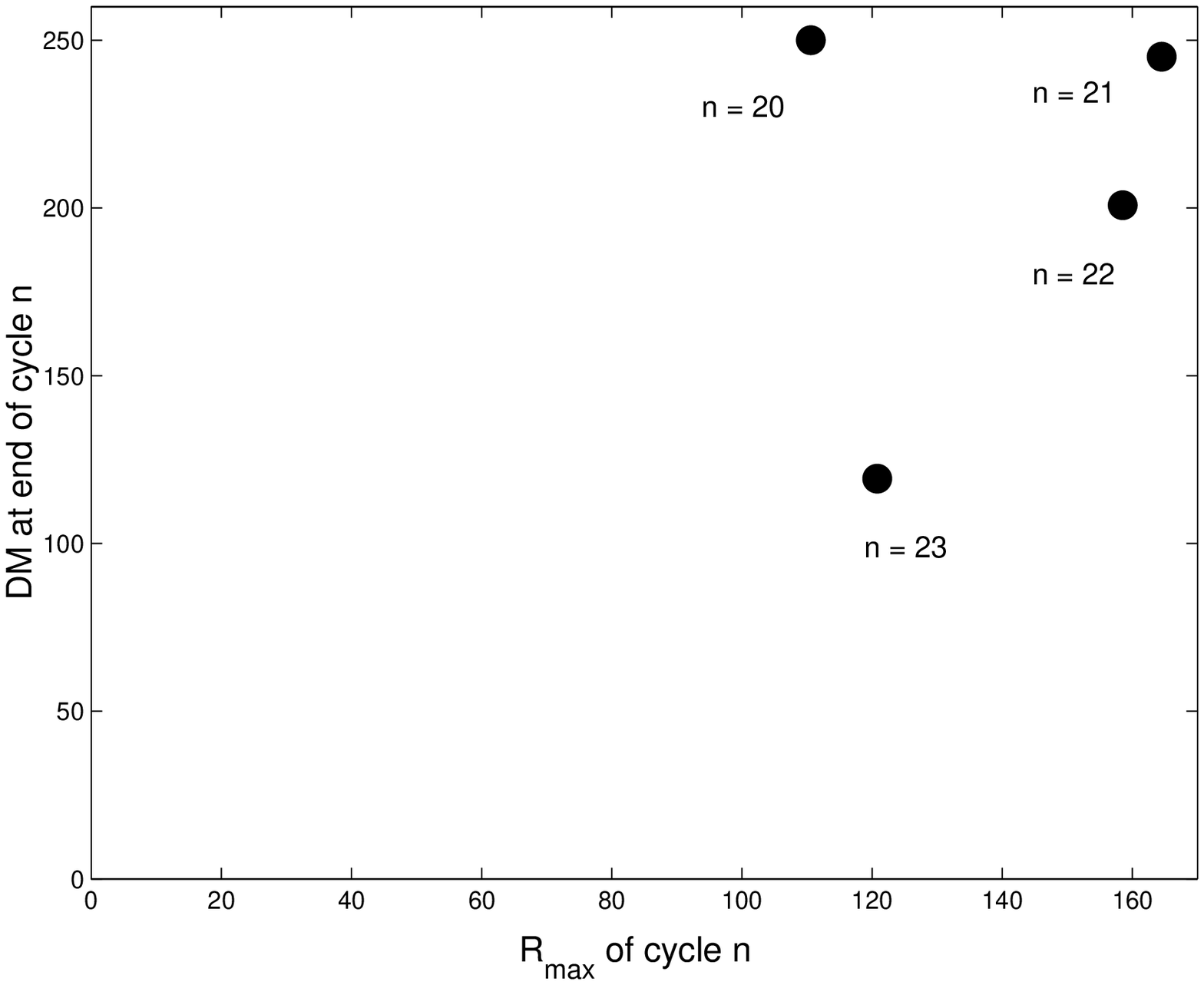}
\end{minipage}
\caption{The left panel shows a plot of the strength
of cycle $n+1$ against the polar field at the end of
cycle $n$. The right panel shows a plot of the polar field
at the end of cycle $n$ against the strength of the cycle
$n$. From Choudhuri [49].}
\end{figure}

If there is really a correlation between the polar field at the
sunspot minimum and the next cycle, then one can use the polar field
to predict the strength of the next cycle [50].  
Since the polar field in the just concluded minimum
has been rather weak (as seen in
Fig.~4), several authors [51, 52] 
suggested that the coming cycle~24 will be rather
weak.  Very surprisingly, the first theoretical prediction based
on a dynamo model made by Dikpati and Gilman [53] is that the
cycle~24 will be very strong.    Dikpati and Gilman [53]
assumed the generation of the poloidal field from the toroidal
field to be deterministic, which is not supported by observational
data shown in the right panel
of Fig.~11. Tobias, Hughes and Weiss [54] make the
following comment on this work:
``Any predictions made with such models should be treated with extreme 
caution (or perhaps disregarded), as they lack solid physical underpinnings.''
While we also consider many aspects of the Dikpati--Gilman work wrong which
will become apparent to the reader soon, we 
cannot also accept the opposite extreme viewpoint
of Tobias, Hughes and Weiss [54], who suggest that the solar dynamo is a
nonlinear chaotic system and predictions are impossible or useless. If that
were the case, then we are left with no explanation for the correlation
seen in the left panel of Fig.~11.  

\begin{figure}
\center
\includegraphics[width=12cm]{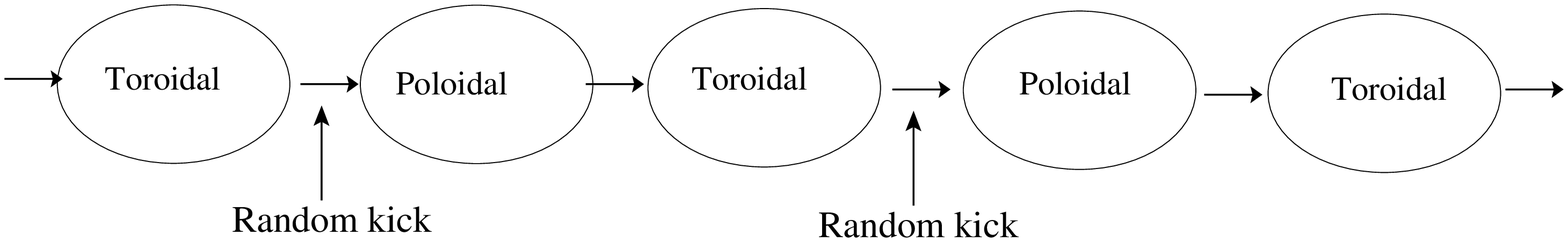}
\caption{A schematic cartoon of the oscillation between toroidal
and poloidal components, indicating the epochs when the system
is subjected to random kicks.}
\end{figure}

Let us now finally come to the
theoretical question as to what produces the variabilities
of cycles and whether we can predict the strength of a cycle before its advent.
Some processes in nature can be predicted and some not.  We can easily
calculate the trajectory of a projectile by using elementary mechanics.
On the other hand, when a dice is thrown, we cannot predict which side
of the dice will face upward when it falls.  Is 
the solar dynamo more like the trajectory 
of a projectile or more like the throw of a dice? Our point of view is that the
solar dynamo is not a simple unified process, but a complex combination of
several processes, some of which are predictable and others not.  Let
us look at the processes which make up the solar dynamo.
 
The flux transport dynamo model combines three basic processes. (i) The
strong toroidal field is produced by the stretching of the poloidal
field by differential rotation in the tachocline. 
(ii) The toroidal field generated in the tachocline
gives rise to sunspots due to magnetic buoyancy and then the decay
of tilted bipolar sunspots produces the poloidal field by the 
Babcock--Leighton mechanism.  (iii) The poloidal field is
advected by the meridional circulation first to high latitudes and then down
to the tachocline, while diffusing as well.   We believe that
the processes (i) and (iii) are reasonably ordered and deterministic.
In contrast, the process (ii) involves an element of randomness due
to the following reason.  The poloidal field produced from the decay of 
a tilted bipolar region by the Babcock--Leighton process depends on
the tilt.  While the average tilt of bipolar regions at a certain
latitude is given by Joy's law, we observationally find quite a large
scatter around this average.
As we already pointed out,  the action
of the Coriolis force on the rising flux tubes gives rise to Joy's
law [27], whereas convective buffeting of 
the rising flux tubes in the upper layers of the convection zone causes the
scatter of the tilt angles [33]. 
This scatter in the tilt angles certainly introduces a
randomness in the generation process of the poloidal field
from the toroidal field.  Choudhuri,
Chatterjee and Jiang [48] identified it as the main source of
irregularity in the dynamo process, which is in agreement 
with Fig.~12. It may be noted that Choudhuri [55] was the
first to suggest several years ago 
that the randomness in the poloidal field
generation process is the source of fluctuations in the dynamo.

\begin{figure}
\center
\includegraphics[width=5cm]{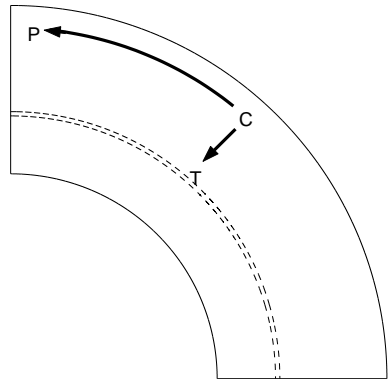}
\caption{A sketch
indicating how the poloidal field produced at  C during a
maximum gives rise to the polar field at P during the
following sunspot minimum and the toroidal field at T during the next
sunspot maximum. From [46].}
\end{figure}

The poloidal field gets built up during the declining phase of
the cycle and becomes concentrated near the poles during the
sunspot minimum.
The polar field at the sunspot minimum produced in a theoretical 
mean field dynamo
model is some kind of `average' polar field during a typical
sunspot minimum.  The observed polar field during
a particular sunspot minimum may be stronger or weaker than this average 
field. The theoretical dynamo model has to be updated by feeding
the information of the observed polar field in an appropriate
way, in order to model actual cycles.
Choudhuri, Chatterjee and Jiang [48] proposed to model this
in the following way. They ran the dynamo code from a
minimum to the next minimum in the usual way.  After stopping
the code at the minimum,
the poloidal field of the theoretical model was multiplied
by a constant factor everywhere above $0.8 R_{\odot}$ to bring
it in agreement with the observed poloidal field.  Since
some of the poloidal field at the bottom of the convection zone may
have been produced in the still earlier cycles, it is left unchanged
by not doing any updating below $0.8 R_{\odot}$.  Only the poloidal
field produced in the last cycle which is concentrated in the upper
layers gets updated. After 
this updating which takes care of the random kick shown
in Fig.~12, we run the code till the next minimum, when the code
is again stopped and the same procedure is repeated. 
Our solutions are now no longer
self-generated solutions from a theoretical model alone, but are
solutions in which the random aspect of the dynamo process has been
corrected by feeding the observational data of polar fields into
the theoretical model.

Before presenting the results obtained with this procedure, we come
to the question how the correlation between the polar field at the sunspot
minimum and the strength of the next cycle as seen in the left panel of Fig.~11 may
arise.  This was first explained by Jiang, Chatterjee and Choudhuri
[46]. The Babcock--Leighton process would first produce the poloidal
field around the region C in Fig.~13.  Then this poloidal field will
be advected to the polar region P by meridional circulation and will
also diffuse to the tachocline T. In the high diffusivity model, this
diffusion will take only about 5 years and the toroidal field of the
next cycle will be produced from the poloidal field that has diffused
to T.  If the poloidal field produced at C is strong, then both the
polar field at P at the end of the cycle and the toroidal field at T
for the next cycle will be strong (and vice versa).  We thus see that
the polar field at the end of a cycle and the strength of the next
cycle will be correlated in the high diffusivity model.  But this
will not happen in the low diffusivity model where it will take more
than 100 years for the poloidal field to diffuse from C to T
and the poloidal field reaches the tachocline only due to the
advection by meridional circulation taking a time of about 20 years. If we
believe that the 3 data points in the left panel of 
Fig.~11 indicate a real correlation,
then we have to accept the high diffusivity model!

\begin{figure}
\center
\includegraphics[width=11cm,height=4.5cm]{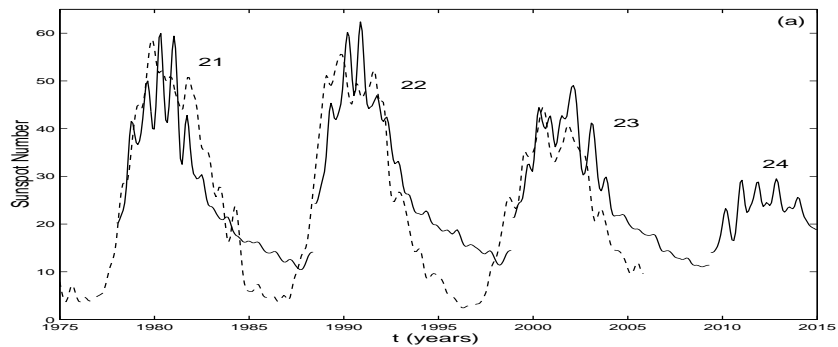}
\caption{The theoretical monthly sunspot number 
(solid line) for the last few years as well as 
the upcoming next cycle, plotted along with the observational data (dashed line)
for the last few years. From Choudhuri, Chatterjee and
Jiang [48].}
\end{figure}

Finally the solid line in
Fig.~14 shows the sunspot number calculated from our high
diffusivity model [48]. Since
systematic polar field measurements are available only from the
mid-1970s, the procedure outlined above could be applied only from
that time.  It is seen from Fig.~14 that our model matches the last
three cycles (dashed line) reasonably well and predicts a weak cycle 24. 
It should be stressed that this is an inevitable consequence of
the high diffusivity model in which the strength of the cycle is
correlated with the polar field in the previous sunspot minimum and we
have fed the information in our calculation that the polar field
in the just-concluded minimum was weak.   
We now wait for the Sun-god to give
a verdict on this prediction within a couple of years. 
It may be mentioned that over the last few years several authors
[43, 46, 56--59] have given several independent arguments in support
of the high diffusivity model. If the next cycle~24 turns out to
be weak (for which there are already enough indications), then
that will provide a further support for the high diffusivity model.

One important related question is whether our dynamo model
can explain occurrences of extreme events like the Maunder 
minimum in the seventeenth century. Choudhuri and Karak [60]
showed that the flux transport dynamo model can reproduce
the Maunder minimum if we introduce a set of assumptions in
the theoretical model.  Whether this set of assumptions necessary
for producing the Maunder minimum is justified on statistical
grounds is an important question which needs to be investigated.

While the fluctuations in the Babcock--Leighton process seem to
be the main source of irregularities in the sunspot cycle, the
meridional circulation also has fluctuations and it has become
apparent in the last few years that the fluctuations in meridional
circulation also introduces irregularities in sunspot cycles [61]. Since
this topic has started being studied systematically only
recently [62--64], it would be premature to provide a summary
of it here.  It seems that the nonlinear aspects
of the equations can also play important roles and there are some
indications that the solar dynamo may be close to a point of chaotic
bifurcation [65, 66]. We are certainly
far from a full theoretical understanding of the irregularities
of the sunspot cycle.
 
\def\apj{{\it Astrophys.\ J.}}
\def\mnras{{\it Mon.\ Notic.\ Roy.\ Astron.\ Soc.}}
\def\sol{{\it Solar Phys.}}
\def\aa{{\it Astron.\ Astrophys.}}
\def\gafd{{\it Geophys.\ Astrophys.\ Fluid Dyn.}}

\end{document}